\documentstyle[12pt]{article}
\begin{document}

%  Greek letters
\def\a{\alpha}
\def\b{\beta}
\def\ch{\chi}
\def\d{\delta}
\def\e{\epsilon}
\def\f{\phi}
\def\g{\gamma}
\def\h{\eta}
\def\i{\iota}
\def\j{\psi}
\def\k{\kappa}
\def\l{\lambda}
\def\m{\mu}
\def\n{\nu}
\def\o{\omega}
\def\p{\pi}
\def\q{\theta}
\def\r{\rho}
\def\s{\sigma}
\def\t{\tau}
\def\u{\upsilon}
\def\x{\xi}
\def\z{\zeta}
\def\D{\Delta}
\def\F{\Phi}
\def\G{\Gamma}
\def\J{\Psi}
\def\L{\Lambda}
\def\O{\Omega}
\def\P{\Pi}
\def\S{\Sigma}
\def\U{\Upsilon}
\def\X{\Xi}
\def\T{\Theta}

\def\Ab{\bar{A}}
\def\gi{g^{-1}}
\def\li{{ 1 \over \l } }
\def\lb{\l^{*}}
\def\zb{\bar{z}}
\def\ub{u^{*}}
\def\vb{v^{*}}
\def\Tb{\bar{T}}
\def\pp {\partial }
\def\pb {\bar{\partial }}
\def\be{\begin{equation}}
\def\ee{\end{equation}}
\def\ben{\begin{eqnarray}}
\def\een{\end{eqnarray}}

\hsize=16.5truecm
\addtolength{\topmargin}{-0.8in}
\addtolength{\textheight}{1in}
%\vsize=26truecm
\hoffset=-.5in

\thispagestyle{empty}
\begin{flushright} \ June \ 1995\\
SNUCTP 95 - 66\\
hep-th/9506087\\
\end{flushright}
\begin{center}
 {\large\bf Duality in Complex sine-Gordon Theory  }\\[.1in]
\vglue .5in
 Q-Han Park\footnote{ E-mail address; qpark@nms.kyunghee.ac.kr }
\\[.2in]
{and}
\\[.2in]
H. J. Shin\footnote{ E-mail address; hjshin@nms.kyunghee.ac.kr }
\\[.2in]
{\it
Department of Physics \\
and \\
Research Institute of Basic Sciences \\
Kyunghee University\\
Seoul, 130-701, Korea}
\\[.2in]
{\bf ABSTRACT}\\[.2in]
\end{center}
New aspects of the complex sine-Gordon theory are addressed through
the reformulation of the theory in terms of the gauged Wess-Zumino-Witten
action. A dual transformation between the theory for the coupling constant
$\b > 0$ and the theory for $\b < 0$ is given which agrees with the
Krammers-Wannier duality in the context of perturbed conformal field theory.
The B\"{a}cklund transform and the nonlinear superposition rule for the
complex sine-Gordon theory are presented and from which, exact solutions,
solitons and breathers with U(1) charge, are derived. We clarify topological
and nontopological nature  of neutral and charged solitons respectively,
and discuss about the duality between the vector and the axial U(1) charges.
\vglue .1in

\newpage
The complex sine-Gordon theory,  which generalizes the well-known
sine-Gordon theory with an internal U(1) degree of freedom, first appeared
as a model of relativistic vortices in a superfluid \cite{lund}, and also
independently in a treatment of O(4) nonlinear sigma model\cite{pohl}.
Recently, Bakas has shown that the complex sine-Gordon theory may be
reformulated in terms of the gauged Wess-Zumino-Witten(WZW) action and
interpreted the theory as the integrably deformed SU(2)/U(1)-coset model
for the parafermions in the large N limit for the level N\cite{bakas}.
This led to subsequent generalizations of the sine-Gordon and the complex
sine-Gordon theories to other coset cases\cite{park}\cite{shin1}\cite{hol1}
as well as generalizations of solitons and breathers with internal degrees of
freedom\cite{shin2}. Reformulation of the theory as the deformed coset
model also provides a natural explanation for the behavior of exact
factorizable S-matrix\cite{hol2}.

In this Letter, we report new aspects of the complex sine-Gordon theory
which arises from the reformulation of the theory in terms of the gauged WZW
action and a specific choice of the gauge described later.
We find the exact duality between
the theories for the coupling constant $\b > 0$ and $\b <0$. We
show that this agrees  precisely with the Krammers-Wannier duality
between the spin variables $s_{j} $ and the dual spin variables $\m_{j}$ of
$Z_{N}$-parafermion theory\cite{zam}.\footnote{The changing sign of coupling
constants $\b $ under the
Krammers-Wannier duality has been pointed out by Bakas and the duality
between the two theories was suggested but without an explicit duality
transform rule\cite{bakas}.} We derive the B\"{a}cklund transform
and the nonlinear superposition rule for the complex sine-Gordon theory
from the gauged WZW action in the gauge $A=\Ab =0$, from which exact
solutions, solitons and breathers with U(1) charge, are obtained. We
clarify the topological nature of soliton solutions. It is shown that
charged solitons are in general nontopological solitons and become topological
only when they become neutral. We also address the duality between the axial
and the vector U(1) charges of the complex sine-Gordon theory.
\\

The complex sine-Gordon theory in terms of the  gauged WZW action is
given by
\ben
I(g,A, \Ab , \b ) &=& I_{WZW}(g) +
{1 \over 2\pi }\int \mbox {Tr} (- A\pb g \gi + \Ab \gi \pp g
 + Ag\Ab \gi - A\Ab ) \nonumber \\
&-& {\b \over 2\pi }\int \mbox{Tr}gT\gi \Tb
\een
where $I_{WZW}(g)$ is the SU(2)-WZW action for a map $g  :
M \rightarrow $ SU(2) on two-dimensional Minkowski space $M$.  The
connection $A, \Ab$ gauge the anomaly free diagonal subgroup U(1) of
SU(2) and $ T = -\Tb = i\s_{3} = \mbox{diag}(i, -i)$ where $\s_{i}; i=1,2,3$
are Pauli matrices.
This action possesses the local U(1) vector symmetry as well
as the global U(1) axial symmetry. The equation of motion of the action (1)
takes a zero curvature form,
\be
\d_{g}I = -{1 \over 2\pi }\int \mbox{Tr}
[\  \pp + \gi \pp g + \gi A g + \b\l T \ , \ \pb + \Ab +
{1 \over \l }\gi \Tb g \ ]\gi \d g  = 0
\ee
which, together with the constraint equation
\ben
\d _{A}I(g,A,\Ab ) &=& {1 \over 2\pi }\int \mbox{Tr} ( \ - \pb g
\gi + g\Ab \gi - \Ab \  )\d A
= 0 \nonumber \\
\d _{\Ab }I(g,A,\Ab ) &=& {1 \over 2\pi }\int \mbox{Tr} ( \  \gi
\pp g  +\gi A g - A \ )\d\Ab = 0 \ ,
\een
defines the theory at the classical level. In particular, the diagonal
part of the equation of motion (2), when combined with the constraint
equation, results in the flatness condition of the connection $A, \Ab$
\be
\pp \Ab - \pb A = 0 .
\ee

In order to show that Eqs.(2) and (3) are indeed the
complex sine-Gordon equation, we introduce a local parametrization
of $g$,
\be
g = \pmatrix{ u & i\sqrt{1-u\ub }e^{ i \q }  \cr
i\sqrt{1-u\ub } e^{ -i \q} & \ub }  \ ,
\ee
and denote that $A=a\s_{3} , \  \Ab = \bar{a} \s_{3}$.
 Then, the constraint equation (3) can be solved for $a,
\bar{a} $ such that
\ben
a &=& {\ub \pp u - u\pp \ub \over 4(1-u\ub )} - {i \over 2}\pp \q
\nonumber  \\
\bar{a} &=& {u\pb \ub - \ub \pb u \over 4(1-u\ub )} - {i \over 2}\pb \q ,
\een
which may be used to bring Eqs.(2) - (4) into a more conventional form of the
complex sine-Gordon equation,
\ben
\pp\pb u + {\ub \pp u \pb u \over 1-u\ub } + 4\b u(1-u\ub ) &=& 0 \nonumber \\
\pp\pb \ub + {u \pp \ub \pb \ub \over 1-u\ub } + 4\b \ub (1-u\ub ) &=& 0 .
\een
After the integration over $A, \Ab$, the action (1) reduces to
\ben
I &=& {1 \over 4\pi }\int \biggl[ {\pp u \pb \ub + \pb u \pp\ub \over 1-u\ub }
-i\pp \q (u\pb \ub - \ub \pb u )-i\pb \q (\ub \pp u -u\pp \ub ) \biggr]
\nonumber \\
& &+ {1 \over 2\pi }\int (\pp u \pb \ub - \pb u\pp \ub ) \ln (-i\sqrt{1-u\ub }
e^{-i \q }) -{\b \over \pi }\int (2u\ub -1) \ .
\een
The second term with a logarithmic expression breaks the reality
condition, whose peculiar behavior arises from the fact that it comes from the
Wess-Zumino term. However, this term does not contribute to the equation of
motion in Eq.(7) but is needed for the current conservation in Eq.(38).
As we see below, the nonreality of the term does not cause any physical
problems.
 The singularity at $u\ub =1 $  comes from the elimination of $A, \Ab$
and is not a real singularity of the theory. At $u\ub =1$,
$A$ and $ \Ab$ become indeterminate from the
constraint equation thereby causing the aparrent sigularity over the
integration. Since $u\ub =1$ is precisely the vacuum of the theory for $\b <0$,
it shows that the WZW actional formulation of the
complex sine-Gordon theory is more natural than the conventional expression
in terms of local fields $u$ and $\ub $ only.  Moreover, we show that it
reveals
some salient features of the theory when we introduce the following gauge
fixing; the U(1) vector gauge transform $g \rightarrow \exp(i\f \s_{3})g
\exp(-i\f \s_{3})$ changes $\q $ in Eq.(5) by $\q + 2\f $ while leaving $u$ and
the complex sine-Gordon equation (7) itself invariant. Owing to Eq.(4), we
may fix the gauge by setting $A = \Ab =0$. This is possible when the holonomy
of the flat connection is trivial which is the case for the flat Minkowski
base manifold. The constraint equation in the gauge $A=\Ab =0$
becomes
\be
{\ub \pp u - u\pp \ub \over 4(1-u\ub )} - {i \over 2}\pp \q = 0  \ , \
 {u\pb \ub - \ub \pb u \over 4(1-u\ub )} - {i \over 2}\pb \q  = 0
\ee
which may be  solved for $\q $ by integration,
\be
\q = {1 \over 2 i }\int dz {\ub \pp u - u\pp \ub \over 1 - u\ub } +
d\zb  {u \pb \ub - \ub \pb u \over 1 - u\ub } .
\ee
In the gauge $A=\Ab=0$, we assert that the theory is symmetric under the
\underline{duality transform};
\be
g \leftrightarrow i\s_{1} g \ , \ \b \leftrightarrow -\b
\ee
which interchanges the sign of the coupling constant $\b $. It is easy to check
that the equation of motion (2) and the constraint equation (3) with $A=\Ab =0$
are both invariant under the duality transform. In components, if we define the
dual
field $g^{D}$ by
\be
g^{D} \equiv \pmatrix{ v & i\sqrt{1-v\bar{v} }e^{ i \f }  \cr
i\sqrt{1-v\bar{v} } e^{ -i \f } & \bar{v} }
= i\s_{1}g =  \pmatrix{ -\sqrt{1-u\ub }e^{ i \q } & i\ub   \cr
iu & -\sqrt{1-u\ub } e^{ i \q}  } ,
\ee
 the duality transform is expressed by
\be
v = -\sqrt{1-u\ub }e^{-i\q }, \ u = \sqrt{1-v\bar{v}}e^{-i\f } \ \mbox{and} \
\b
\leftrightarrow  -\b \ .
\ee
The theory is also symmetric under another type of duality transform,
\be
g \leftrightarrow ig\s_{1}  \ , \ \b \leftrightarrow -\b
\ee
which is identical with Eq.(11) except for the change of the sign of $\q $.
The duality in Eq.(11) may be compared with the Krammers-Wannier duality of
the $Z_{N}$ parafermion theory bewteen the spin variables $s_{j} $ and the
dual spin variables $\m_{j}$.
These spin variables are given by $s_{k} = \phi^{(k)}_{k,k}$ and $\m_{k} =
\phi^{(k)}_{-k,k}$ where $\phi^{(2j)}_{2m , 2\bar{m}} ( - j \ge m, \bar{m}
\ge j  \ \ ; \   j = 0, 1/2,\cdots , N/2 )$ are the primary
fields of the $Z_{N}$ parafermion theory which is related to those of the
WZW theory by
\be
\Phi^{(j) }_{m , \bar{m}}(z, \zb ) = \phi^{(2j)}_{2m , 2\bar{m}}(z, \zb )
:\exp\biggl[  {im \over \sqrt{N}}\ch (z) +  {i\bar{m} \over \sqrt{N}}\bar{\ch }
(\zb ) \biggr] : .
\ee
where $\ch (z) ,\bar{\ch} (\zb )$ are U(1) scalar
fields\cite{kniz}\cite{gepner}.
In the case $j=1$, the primary fields of WZW theory $\Phi^{(1)}$ may be given
by the adjoint representation of $g$,
\be
\Phi^{(1)}_{ab} = \mbox{Tr} \gi T_{a}g T_b
\ee
where $T_{a}$ are generators of the algebra $su(2)$.  In the limit
$N \rightarrow \infty $, Eq.(15) shows that the spin variables can be written
directly in terms of the WZW primary fields such that
\ben
s_{2} &=& \Phi^{1}_{1,1} = \mbox{Tr} \gi T_{+} g T_{+} = (1-u\ub )e^{-2i\q }
\nonumber \\
\m_{2} &=& \Phi^{1}_{-1,1} = \mbox{Tr} \gi T_{-} g T_{+} = u^2
\een
where $T_\pm \equiv (\s_{1} \pm i\s_{2} )/2$. Thus, the Krammers-Wannier
duality
which interchange $s_{2}$ and $\m_{2}$ agrees precisely with the duality
transform in Eq.(13).
If we assume that $u =\ub = \cos \varphi$, then the complex
 sine-Gordon equation reduces to the well-known sine-Gordon equation
$\pp\pb \varphi = 2\b \sin 2\varphi $ and the duality transform reduces to the
more
familar one,
\be
\cos \varphi \leftrightarrow \sin \varphi \ \ \mbox{or} \ \ \varphi
\leftrightarrow
\varphi + (n+{1 \over 2})\pi , \ n \in Z \ ; \ \b \leftrightarrow -\b .
\ee
The duality transform also relates exact solutions of the $\b < 0$ case
with those of the $\b > 0$ case. To better understand the vacuum structure
of the theory, we introduce another parametrization of $g$ which is equivalent
to Eq.(5),
\be
g=e^{i\eta \s_{3}}e^{i\varphi (\cos{\q }\s_{1} + \sin{\q }\s_{2})}e^{i\eta
\s_{3}}
= \pmatrix{ e^{2i\eta }\cos{\varphi } & i\sin{\varphi }e^{-i\q } \cr
i\sin{\varphi }e^{i\q } & e^{-2i\eta }\cos{\varphi } }\ .
\ee
The potential term then becomes
\be
V = {\b \over \pi }\cos{2\varphi }
\ee
so that the vacuua  for $\b < 0$ are specified by
\be
\varphi = n\pi , \ n \in Z  \ ; \
g_{\mbox{vac}} = \pmatrix{\pm e^{i\O } & 0 \cr 0 & \pm  e^{-i\O } } ,
\ee
while those for $\b > 0$ are
\be
\varphi = (n+{1\over 2})\pi , \ n \in Z \ ; \
g_{\mbox{vac}} = \pmatrix{ 0 &\pm  ie^{-i\O } \cr \pm ie^{i\O } & 0 } \ .
\ee
It is easy to check that these two vacuum solutions are related by the
duality transform. In order to adjust the vacuum to possess zero potential
energy, one has to add a constant term to the potential in which case, the
coupling constant $\b $ can be treated as a mass parameter in the perturbative
approach.
The degeneracy of the vacuua predicts the existence of
a soliton solution
with a topological soliton number $\D n = n_{1}-n_{2}$ which interpolates
two different vacuua characterized by integers $n_{1}$ and $n_{2}$.
In the following, we derive solitons explicitly from the B\"{a}cklund transform
for
the complex sine-Gordon theory. The B\"{a}cklund transform can be
compactly given by
\be
\J_{g}(\l ) = {\l \over \l - i\d / \sqrt{|\b | }}\biggl( 1+ {\d \over
\sqrt{|\b | }\l } \gi \Tb f \biggr) \J_{f} (\l )  \ ,
\ee
where $\d$ is an arbitrary parameter and $\J_{g} , \J_{f}$
satisfy the linear equations,
\ben
(\pp + \gi \pp g + \b \l T)\J_{g} = 0 \ &,& \
(\pb + {1\over \l}\gi \Tb g )\J_{g} = 0 \nonumber \\
(\pp + f^{-1} \pp f + \b \l T)\J_{f} = 0 \ &,& \
(\pb + {1\over \l}f^{-1} \Tb f )\J_{f} = 0 .
\een
If we eliminate $\J_{g} , \J_{f}$ from Eqs.(23) and (24), we obtain the
B\"{a}cklund
transform in terms of $g$ and $f$ only,
\ben
\gi \pp g - f^{-1}\pp f - {\d \b \over  \sqrt{|\b | }}[\ \gi \Tb f \ ,
\ T \ ] &=& 0  \nonumber \\
\pb g \gi \Tb - \Tb \pb f f^{-1} + {\sqrt{ | \b | }\over \d } ( gf^{-1}\Tb -
\Tb g f^{-1} )&=& 0 \ .
\een
If we apply $\pb $ to the first equation in Eq.(25) and combine with the second
equation,  we see that  $g$ and $f$ both satisfy the complex sine-Gordon
equation. It is easy to convince that
Eq.(25) is also consistent with the constraint equation (3) for $A=\Ab=0$.
 Thus, the B\"{a}cklund transform generates a new solution $g$ from a
known solution $f$ via the first order differential equation (25). In
particular,
1-soliton solution can be obtained by applying the B\"{a}cklund transform  to
the
vacuum solutions in Eqs.(21) and (22).
For $\b < 0$,  Eq.(25),  with $f = g_{\mbox{vac}}$ in Eq.(21), becomes in
components
\ben
\pp u + 2\d \sqrt{-\b }e^{i \O }(1-u\ub ) &=& 0 \nonumber \\
\pb u - { 2  \sqrt{-\b }\over \d }e^{i \O }(1-u\ub ) &=& 0
\een
and their complex conjugates. These equations and Eq.(10) may be readily
integrated to yield the 1-soliton solution,
\ben
u &=& -e^{i\O }\cos(\a ) \tanh [2\sqrt{-\b }
\cos(\a ) { x - Vt \over \sqrt{1-V^{2}}} ] - ie^{i\O }\sin (\a )
\nonumber \\
\q &=& -2\sqrt{-\b}\sin ( \a ) { t-Vx \over \sqrt{1 -V^{2}}} \ .
\een
where  $V \equiv (1-\d^{2})/(1+\d^{2})$ is the velocity of the soliton,
$\cos{\a } $ and $ \sin{\a }; -\pi \le \a < \pi $ are the constants of
integration and $ t \equiv {1 \over 2}(z + \zb ) ,
\ x \equiv {1 \over 2}(z - \zb ) $.
Similarly, we obtain the B\"{a}cklund equation for $\b > 0$ from the vacuum
in Eq.(22),
\ben
\pp u - 2\d \sqrt{\b }e^{i(\q +  \O ) }u\sqrt{1-u\ub }  &=& 0 \nonumber \\
\pb u + { 2  \sqrt{\b }\over \d }e^{-i (\q +  \O ) }u\sqrt{1-u\ub } &=& 0 \ ,
\een
from which we obtain the 1-soliton for $\b > 0$,
\ben
u &=& { \cos(\a ) \over \cosh [2\sqrt{\b}\cos(\a )
{ x-Vt \over \sqrt{1-V^{2}}}] } \exp ( 2i\sqrt{\b} \sin(\a )
{t - Vx \over \sqrt{1-V^{2}}} )
\nonumber \\
\q &=& -\O - \tan^{-1}( \tan (\a ) \coth [2\sqrt{\b } \cos( \a ) {x -Vt \over
\sqrt{1-V^{2}}}]) \ .
\een
This agrees with the 1-soliton for $\b < 0$ through the duality transform.
The $u$-part of the soliton solutions are obtained previously by Lund and
Regge\cite{lund}
 for $\b <0$ and by Getmanov\cite{getmanov} for $\b > 0$.
Before discussing the physical meaning of the 1-soliton solutions, we note
that 2-solitons and breather solutions can be also obtained from the following
nonlinear superposition rule; let $g_{1}$ and $g_{2}$ be two solutions
generated by  the B\"{a}cklund transform from a known
solution $g_{0}$ with respect to the B\"{a}cklund parameters $\d_{1}$
and $\d_{2}$. If we take the B\"{a}cklund transform again on $g_{2}$ with
the parameter $\d_{1}$ and also on $g_{1}$ with $\d_{2}$ and require the
commutability of the process, i.e. require that
both of them result in the same $g_{3}$, then we have from Eq.(23)
\ben
\J_{g_{3}} &=& {\l \over \l - i\d_{2} / \sqrt{|\b | }}
{\l \over \l - i\d_{1} / \sqrt{|\b | }}
( 1+ {\d_{2} \over \sqrt{| \b | }\l } \gi_{3} \Tb g_{1} )
( 1+ {\d_{1} \over
 \sqrt{| \b | }\l }\gi_{1} \Tb g_{0} ) \J_{g_{0}} \nonumber \\
 & = &
{\l \over \l - i\d_{1} / \sqrt{|\b | }}{\l \over \l - i\d_{2} / \sqrt{|\b | }}
( 1+ {\d_{1} \over \sqrt{|\b | }\l } \gi_{3} \Tb g_{2} )
( 1+ {\d_{2} \over \sqrt{|\b | }\l }
\gi_{2} \Tb g_{0} ) \J_{g_{0}} \ .
\een
This may be solved for $g_{3}$ in terms of $g_{1}$ and $g_{2}$ such that
\be
g_{3} = \Tb (\d_{1}g_{2} -\d_{2}g_{1} )g_{0}^{-1}\Tb^{-1}
(\d_{1}\gi_{1} - \d_{2}g_{2}^{-1} )^{-1} \ ,
\ee
which represents a nonlinear superposition rule for the complex sine-Gordon
theory.
In particular, if we choose $g_{0}$ as the vacuum solution in Eq.(21) then
$g_{1}$ and $g_{2}$ become
\ben
g_{k} &=& \pmatrix{ u_{k}& i\sqrt{1-u_{k}\ub_{k}}e^{i\q_{k}}
 \cr i\sqrt{1-u_{k}\ub_{k}}e^{-i\q_{k}} & \ub_{k} }  \nonumber \\
u_{k} &=& -e^{i\O }(\cos (\a_{k})\tanh [2\sqrt{-\b }\cos (\a_{k}) \S_{k}]
 + i\sin (\a_{k} )) \nonumber \\
\ \q_{k} &=& -2\sqrt{-\b }\sin (\a_{k})\T_{k} \ ; \ k = 1,2
\een
whereas $g_{3}$ is given by Eq.(31). $\S_{k}$ and $\T_{k}$ are to be determined
by
the choice of parameters $\d_{1}$ and $\d_{2}$. Following three specific
choices of
parameters are of interest;
\\
\underline{ i) soliton-soliton scattering};
\be
 \d_{1} = -{1 \over \d_{2}} =  \sqrt{ 1-V \over 1+V} , \
\S_{1} = { x - Vt \over \sqrt{1-V^{2}}}  , \
\T_{1} = { t - Vx \over \sqrt{1-V^{2}}}   \ ; \
\S_{2} = { -x - Vt \over \sqrt{1-V^{2}}}  , \
\T_{2} = { -t - Vx \over \sqrt{1-V^{2}}}
\ee
\underline{ii) soliton-antisoliton scattering}
\be
 \d_{1} = {1 \over \d_{2}} =  \sqrt{ 1-V \over 1+V} , \
\S_{1} = { x - Vt \over \sqrt{1-V^{2}}}  , \
\T_{1} = { t - Vx \over \sqrt{1-V^{2}}}   \ ; \
\S_{2} = { x+Vt  \over \sqrt{1-V^{2}}}  , \
\T_{2} = { t+Vx \over \sqrt{1-V^{2}}}
\ee
\underline{iii) breather }
\be
 \d_{1} = {1 \over \d_{2}} =  \sqrt{ 1-iV \over 1+iV} , \
\S_{1} = { x-iVt \over \sqrt{1+V^{2}}}  , \
\T_{1} = { t-iVx \over \sqrt{1+V^{2}}}   \ ; \
\S_{2} = { x+iVt \over \sqrt{1+V^{2}}}  , \
\T_{2} = { t+iVx \over \sqrt{1+V^{2}}}
\ee
which can be shown to agree with the sine-Gordon cases when $\a_{k} = 0$.

Finally, we address the physical meaning of the parameter $\a $ which can be
inferred from the global U(1) symmetries of the theory.
Recall that the complex sine-Gordon theory as defined
in Eq.(1) is invariant under the local vector U(1) and the global axial U(1)
transformations. After the gauge fixing $A=\Ab =0$, the residual symmetries
are the global vector and axial U(1). The global axial vector transformation is
\be
g \rightarrow e^{i\e \s_{3}}ge^{i\e \s_{3}}
\ee
or, in components,
\be
u \rightarrow ue^{2i\e } \ , \ \ub \rightarrow  \ub e^{-2i\e } \ , \ \q
\rightarrow \q \ .
\ee
The conserved axial currents $J_{A} , \bar{J}_{A} $ satisfying the relation, $
\pb J_{A} +
\pp \bar{J}_{A} = 0$, can be obtained directly from the action in Eq.(8) by the
Noether method,
\ben
J_{A} &=& i {\ub \pp u - u\pp \ub \over 1 - u\ub } + 2\pp (u\ub \q ) + i\pp
F(u\ub )
 \nonumber \\
\bar{J}_{A} &=& i {\ub \pb u - u\pb \ub \over 1 - u\ub } - 2\pb (u\ub \q )
- i\pb F(u\ub ) ,
\een
where the function $F$ satisfies $F^{'}(x) = 2\mbox{ln}(-i\sqrt{1-x} )$.
We caution that the action in Eq.(8) is an effective action where the
constraint Eq.(3) is
imposed. The Noether method can be applied consistently only when the
constraint is
invariant under the axial transformation which is indeed the case. The
conserved
axial charge arising from currents in Eq.(38) is
\be
Q_{A} = -i\int_{-\infty}^{\infty} dx {\ub \pp_{t}u - u\pp_{t}\ub \over 1-u\ub }
-\biggl[ 2u\ub \q +iF(u\ub )\biggr]^{\infty}_{-\infty} \ .
\ee
Thus the axial charges of 1-soliton solutions computed in the rest frame are
\be
Q_{A}(\b < 0 ) = 0 \ ; \ Q_{A}(\b > 0) = 4( \mbox{sign}[\a ]{\pi \over 2} - \a
)
\mbox{ for } \sin{\a } \ne 0
\ee
which shows that the parameter $\a $ characterizes the U(1) charge of
solitons. If $\sin{\a }= 0$, the axial charge is indeterminate because of the
singular behavior of $Q_{A}$ at $u\ub =1$. As stated earlier, this relects
the fact that $A, \Ab$ are indeterminate at $u\ub =1$. Here, we simply
take $Q_{A}$ to be zero. This may be done more rigorously by taking an
appropriate limiting procedure in the gauge $A=\Ab = 0$, which however is
physically irrelevant.
In the case of the vector U(1) symmetry, we note that the duality transform in
Eq.(11)
 relates the axial symmetry with the vector symmetry in the following way; if
we
perform the axial transform in Eq.(36), then the dual field $g^{D} = i\s_{1}g$
changes
according to the vector transform,
\be
g^{D} \rightarrow e^{-i\e \s_{3}}g^{D}e^{i\e \s_{3}} \ .
\ee
This implies that the conserved vector currents and charges are dual to the
axial
ones. That is, the vector charges $Q_{V}$ are given by
\be
Q_{V}(\b < 0) = Q_{A}(\b > 0)  \ ; \
Q_{V}(\b > 0) = Q_{A}(\b < 0)
\ee
whereas the vector currents $J_{V} , \bar{J}_{V}$ can be obtained by taking the
dual transform of the axial currents such that
\ben
J_{V} &=& \pp ( iu\ub \ln {\ub \over u} + iF ) \nonumber \\
\bar{J}_{V} &=& - \pb ( iu\ub \ln {\ub \over u} + iF ) \ .
\een
Another aspect of the parameter $\a$ is that it dictates the topological nature
of
solitons. If $\cos{\a }  = 0 $, the solutions in Eqs.(27) and (29) trivially
reduce
 to the vacuum solutions in Eqs.(21) and (22). If $\sin{\a } =0$, it is easy to
check
that the analyticity of the solution  requires that $\D n = \pm 1$, i.e. it
becomes
a topological 1-soliton (antisoliton).
For $\cos{\a } \ne 0 \mbox{ and } \sin{\a } \ne 0$, the continuity of the
solution
requires that $\D n = 0$, which shows that charged solitons are nontopological.
Nevertheless, they are stable against the decay into multi-nontopological
solitons and
mesons. This semiclassical stability may be shown by the mass
formula\footnote{ Here, the mass constant $m$ is our
coupling constant $2\sqrt{|\b|}$. The constant $\l^2 $ gets renormalized
in the quantum case which in our classical consideration has been set to one.}
 for the nontopological soliton of charge $Q$\cite{hol2}
\be
M(Q) = {4m \over \l^2 }|\sin{ \l^2 Q \over 4 }|, \ |Q| \le {2\pi \over \l^2 } \
{}.
\ee
It is easy to check that $
M(Q_{1}) + M(Q_{2}) > M(Q_{1} + Q_{2}) $ which prohibits the nontopological
soliton
from decaying in two solitons conserving the charge. Also, the decay into the
elementary meson of charge 1 and mass $m$ is suppressed by the bound
$ M(1) = { 4m \over \l^2 }\sin{ \l^{2} \over 4} < m$.

\vglue .2in
{\bf ACKNOWLEDGEMENT}
\vglue .2in
We would like to thank I.Bakas, R. Sasaki and T.Hollowood for discussion.
This work was supported in part by the program of Basic Science Research,
Ministry of Education BSRI-95, and by Korea Science and Engineering
Foundation through CTP/SNU and the Korea-Japan Cooperative Science Program.
\vglue .2in

\end{document}